\documentstyle[amsfonts,amsmath,12pt,graphics]{elsart}
\def\text#1{\mbox{#1}}

\begin{document}

\begin{frontmatter}
\title{ Vortices in a nonminimal Maxwell-Chern-Simons O(3) Sigma Model}

\author[UFC,UECE]{F. S. A. Cavalcante},
\author[UFC,UECE]{ M. S. Cunha} and
\author[UFC]{C. A. S. Almeida\thanksref{e-mail}}
\address[UFC]{ Universidade Federal do Cear\'{a} - Departamento de
F\'{\i}sica \\ C.P. 6030, 60470-455 Fortaleza-Ce, Brazil}
\address[UECE]{ Universidade Estadual do Cear\'{a} - Departamento de
F\'{\i}sica e Qu\'{\i}mica \\Av. Paranjana, 1700, 60740-000
Fortaleza-Ce, Brazil}
\thanks[e-mail]{Electronic address: carlos@fisica.ufc.br}

\begin{abstract}
In this work we consider an Abelian O(3) sigma model coupled
nonminimally with a gauge field governed by Maxwell and
Chern-Simons terms. Bogomol'nyi equations are constructed for a
{\it {specific form}} of the potential and generic nonminimal
coupling constant. Furthermore, topological and nontopological
self-dual soliton solutions are obtained for a critical value of
the nonminimal coupling constant. Some particular static vortex
solutions (topological and nontopological) satisfying the
Bogomol'nyi bound are numerically solved and presented.
\end{abstract}
\end{frontmatter}

\section{INTRODUCTION}

The gauging of a non-linear sigma model, giving rise to a coupling between
scalar (coordinates of the target space) and gauge fields has attracted
interest from different areas. In particular, soliton solutions of the
gauged O(3) Chern-Simons model may be relevant in planar condensed matter
systems \cite{zee,laughlin,polyakov}. Recently, self-dual solutions were
obtained in a (2+1) dimensions gauged O(3) sigma model \cite{schroers}. In
the context of Skyrme models the requirement of gauge symmetry was used to
break the scale invariance of the O(3) sigma model with gauge field dynamics
governed by a Maxwell term, where the gauge group U(1) is a subgroup of
O(3), giving the so called baby Skyrme model \cite{gladi1,piette} (aided by
a potential term). Other works, in which the gauge field dynamics is
governed by a Chern-Simons term, were studied in a series of papers \cite
{kimm,gladi2,arthur}. These systems becomes self-dual for a specific choice
of the Higgs potential and topological as well as nontopological solutions
are present \cite{ghosh1}.

Stern \cite{stern} was the first one to suggest a nonminimal term in the
context of the Maxwell-Chern-Simons electrodynamics intending to mimic an
anyonic behavior without a pure Chern-Simons limit. This term could be
interpreted as a generalization of the Pauli coupling, {\it i. e. }, an
anomalous magnetic moment. It is a specific feature of (2+1) dimensions that
the Pauli coupling exists not only for spinning particles but for scalar
ones too \cite{kogan,lat,torres,georgelin}. As a fundamental result, Stern
showed that, for a particular value of the nonminimal coupling constant, the
field equations of his model have the same form of the field equations of a
{\it pure} Chern-Simons theory minimally coupled.

In this work we consider an {\it Abelian} gauged O(3) sigma model with a
nonminimal coupling, where we have included Maxwell and Chern-Simons terms.
Self-dual soliton solutions are obtained for this model at a critical value
for the nonminimal coupling constant. Our soliton solutions are charged, but
the magnetic flux is topologically quantized only for topological solitons
(where stability is guaranteed by topological arguments). It is important to
emphasize that the Bogomol'nyi bound \cite{bogo} will be achieved with the
introduction of a neutral scalar field. Previously, a {\it non-Abelian}
gauged sigma model with anomalous magnetic moment was discussed \cite{ghosh}%
. However that work did not considered the Higgs potential.

Finally, we give the asymptotic behaviour of the topological and the
nontopological solitons and present the numerically calculated soliton
profiles for {\it both} cases.

\section{THE NONMINIMAL GAUGED O(3) SIGMA MODEL}

In the O(3) sigma model the scalar field $\pmb {\phi }$ is a map from the
(2+1)-dimensional Minkowski space to the two-sphere of unit radius denoted
by $S^2$. In other words, $\pmb {\phi }$ is a three component vector
satisfying the constraint $\pmb {\phi }$ $\cdot $ $\pmb {\phi }=\phi
_1^2+\phi _2^2+\phi _3^2=1$. Now we look for a Lagrangian invariant under
global iso-rotations of the field $\pmb {\phi }$ about a fixed axis ${\bf n}$
$\in S^2$. In order to gauge this symmetry, we choose ${\bf n}=(0,0,1)$ and
select the local U(1) subgroup of the O(3) rotational group.

Therefore, the nonminimal gauged O(3) sigma model is defined by the
following Lagrangian density
\begin{eqnarray}
L &=&-\frac 14F^{\mu \nu }F_{\mu \nu }+\frac \kappa 4\varepsilon ^{\mu \nu
\alpha }A_\mu F_{\nu \alpha }+\frac 12\partial _\mu M\partial ^\mu
M-g\partial _\mu M\partial ^\mu ({\bf n\cdot }\pmb {\phi })  \nonumber \\
&&+\frac 12\nabla _\mu \pmb {\phi }\cdot \nabla ^\mu \pmb {\phi }-U(M,{\bf {n%
}}\cdot \pmb {\phi }),  \label{one}
\end{eqnarray}
where the second term is the called Chern-Simons term and the minimal gauge
covariant derivative $D_\mu \pmb {\phi }\equiv \partial _\mu \pmb {\phi }%
+\left[ eA_\mu \right] {\bf {n}}\times \pmb {\phi }$ was changed by
\begin{equation}
\nabla _\mu \pmb {\phi }=\partial _\mu \pmb {\phi }+\left[ eA_\mu +\frac
g2\varepsilon _{\mu \nu \alpha }F^{\nu \alpha }\right] {\bf {n}}\times \pmb {%
\phi \quad \quad .}  \label{two}
\end{equation}
The real scalar field $M$ is introduced for convenience, as explained later.
This Lagrangian is clearly gauge invariant since the potential depends only
on $\phi _3,$ which is gauge invariant.

Notice that our metric is $\eta _{\mu \nu }=diag(+1,-1,-1)$ with $%
\varepsilon ^{012}=1$ ; $\mu ,\nu =0,1,2$ and $i,j=1,2.$

The equations of motion for the gauge field follows from (\ref{one}) as
\begin{equation}
e{\bf {n}}\cdot \pmb {\phi }\times \nabla ^\mu \pmb {\phi }+\kappa F^\mu
-\varepsilon ^{\mu \nu \alpha }\partial _\alpha \left( g{\bf {n}}\cdot \pmb {%
\phi }\times \nabla _\nu \pmb {\phi }-F_\nu \right) =0  \label{three}
\end{equation}
where $F^\mu \equiv \frac 12\varepsilon ^{\mu \nu \alpha }F_{\nu \alpha }$
is the dual to the field strength and the matter current is given by
\begin{equation}
J^\mu =-e{\bf {n}}\cdot \pmb {\phi }\times \nabla ^\mu \pmb {\phi }\qquad .
\label{four}
\end{equation}
So, equation (\ref{three}) can be rewritten as
\begin{equation}
\partial _\mu \left[ \varepsilon ^{\mu \alpha \nu }\left( \frac geJ_\nu
+F_\nu \right) \right] =J^\alpha -\kappa F^\alpha \qquad .  \label{five}
\end{equation}
The solutions of the first order differential equation
\begin{equation}
F^\alpha =\frac 1\kappa J^\alpha   \label{six}
\end{equation}
are also solutions of (\ref{five}) provided that
\begin{equation}
g=-\frac e\kappa   \label{seven}
\end{equation}
holds.

This result is completely analogous to that presented by Stern \cite{stern}
and later by Torres \cite{torres} in the context of Maxwell-Chern-Simons
electrodynamics.

It is worth mentioning{\it \ }that the $\alpha =0$ component of the equation
(\ref{six}) is the corresponding {\it Gauss law, }namely
\begin{equation}
J_0+\kappa B+\partial _iE_i+\frac ge\varepsilon _{ij}\partial _iJ_j=0
\label{gauss}
\end{equation}
where $B$ and $E_{i\mbox{ }}$ are the magnetic and electric fields
respectively.

Integration of (\ref{gauss}) in the whole space leads to
\begin{equation}
\Phi \equiv \displaystyle\int d^2xB=-\frac q\kappa {\ .}  \label{seven1}
\end{equation}
Therefore, charged vortices with charge $q$ are also tubes of magnetic flux.

Now we construct the energy functional. In this direction, a matter field
equation of motion is required and it is possible to express it in terms of
currents. So, let us consider the following equation of motion for the
scalar field
\begin{equation}
\nabla _\mu \nabla ^\mu \pmb {\phi }=-\frac{\partial V}{\partial \pmb {\phi }%
}\mbox{ .}  \label{eight}
\end{equation}
We define a new matter current ( which is a vector in the target space) as
\begin{equation}
\stackrel{}{{\bf j}^\mu }=-e\pmb {\phi }\times \nabla ^\mu \pmb {\phi \ .}
\label{nine}
\end{equation}
Thus the equation (\ref{four}) can be rewritten in the following form
\begin{equation}
J^\mu =\stackrel{}{{\bf {n}\cdot }}{\bf j}^\mu \qquad .  \label{ten}
\end{equation}
Now the equation of motion (\ref{eight}) turns to
\begin{equation}
\nabla ^\mu J_\mu =e\pmb {\phi }\times \frac{\partial V}{\partial \pmb {\phi
}}\qquad ,  \label{eleven}
\end{equation}
where we have used (\ref{nine}) and (\ref{ten}). For the sake of
completeness it is interesting to express the current $J^\mu $ in terms of a
current without explicit magnetic moment contribution, denoted by $k^\mu
\equiv -e{\bf {n}}\cdot \pmb {\phi }\times $ $D^\mu \pmb {\phi }$. So
\begin{equation}
k^\mu =J^\mu \left[ 1+\left( g{\bf {n}}\times \pmb {\phi }\right) ^2\right] %
\mbox{ .}  \label{twelve}
\end{equation}
Later the expression above will be useful to write the energy functional
properly.

In addition, a gauge invariant topological current $J_{top}^\mu $ can be
defined as
\begin{equation}
J_{top}^\mu =\frac 1{8\pi }\varepsilon ^{\mu \nu \alpha }\pmb {\phi }\cdot
\left[ D_\nu \pmb {\phi }\times D_\alpha \pmb {\phi }-\frac e2F_{\nu \alpha
}\left( 1-{\bf {n}}\cdot \pmb {\phi }\right) \pmb {\phi }\right] {\ ,}
\label{topcurr}
\end{equation}
which leads to a topological charge in the form
\begin{equation}
Q_{top}=\int d^2xJ_{top}^0=\frac 1{4\pi }\int d^2x\left[ \pmb {\phi }\cdot
D_1\pmb {\phi }\times D_2\pmb {\phi }+\frac{eB}2\left( 1-{\bf {n}}\cdot \pmb
{\phi }\right) \right] {\ .}  \label{topcharg}
\end{equation}

\section{SELF-DUAL EQUATIONS}

Usually, the energy functional is obtained by the integral over the $T_{00}$
component of the energy momentum-tensor, which can be obtained by coupling
the fields to gravity and then varying the action with respect to the
metric. Taking into account that the Chern-Simons term does not contribute
because of its metric independence, we have
\begin{eqnarray}
T_{\mu \nu } &=&G(g,\pmb {\phi }{\bf )}F_\mu ^{\;\alpha }F_{\alpha \nu
}+D_\mu \pmb {\phi }\cdot D_\nu \pmb {\phi }+\partial _\mu M\partial _\nu M
\nonumber \\
&&-\frac 12\left[ \partial _\mu M\partial _\nu ({\bf n}\cdot \pmb {\phi }%
)+\partial _\nu M\partial _\mu ({\bf n}\cdot \pmb {\phi })\right] -\eta
_{\mu \nu }L\mbox{ ,}  \label{emt}
\end{eqnarray}
where $G(g,\pmb {\phi }{\bf )}=1-g^2\left( {\bf n}\times \pmb
{\phi }\right) ^2$.

So, the $T_{00}$ component of the energy-momentum tensor can be written as
\begin{eqnarray}
T_{00} &=&\frac 12G\left( B^2+E^2\right) +\frac 12\partial _0M\partial
_0M+\frac 12\partial _iM\partial _iM-g\partial _0M\partial _0({\bf n}\cdot %
\pmb {\phi })  \nonumber \\
&&-g\partial _iM\partial _i({\bf n}\cdot \pmb {\phi })+\frac 12D_0\pmb {\phi
}\cdot D_0\pmb {\phi }+\frac 12D_i\pmb {\phi }\cdot D_i\pmb {\phi }{\bf +}%
U(M,{\bf {n}}\cdot \pmb {\phi })\mbox{ .}  \label{emt0}
\end{eqnarray}
If we use the equations (\ref{two}) and (\ref{twelve}) we obtain the
following expression
\begin{eqnarray*}
\frac 12D_i\pmb {\phi }\cdot D_i\pmb {\phi } &=&-\frac 12g^2\left( {\bf n}%
\times \pmb {\phi }\right) ^2E^2+\frac geF_ik_i\mp gE_i\partial _i\left(
{\bf n}\cdot \pmb {\phi }\right) \\
&&+\frac 12\left( \nabla _1\pmb {\phi }{\bf \pm }\pmb {\phi }\times \nabla _2%
\pmb {\phi }\right) ^2\pm \pmb {\phi }\cdot D_1\pmb {\phi }\times D_2\pmb {%
\phi }\mbox{ .}
\end{eqnarray*}
Then the total energy ${\cal E}$ can be expressed as
\begin{eqnarray}
{\cal E} &=&\int d^2x\left\{ \frac 12G\left[ B\mp \frac eG\left( (1-{\bf n}%
\cdot \pmb {\phi }{\bf )+}\frac \kappa eM+g\left( {\bf n}\times \pmb {\phi }%
\right) ^2M\right) \right] ^2\right.  \nonumber \\
&&\pm eB\left[ (1-{\bf n}\cdot \pmb {\phi }{\bf )+}\frac \kappa eM+g\left(
{\bf n}\times \pmb {\phi }\right) ^2M\right] +\frac 12\left( E_i\pm \partial
_iM\right) ^2  \nonumber \\
&&\mp E_i\partial _iM-\frac 12g^2\left( {\bf n}\times \pmb {\phi }\right)
^2E^2+\frac 12\partial _0M\partial _0M-\frac 12g\partial _0M\partial _0({\bf %
n}\cdot \pmb {\phi })  \nonumber \\
&&-g\partial _iM\partial _i({\bf n}\cdot \pmb {\phi })+\frac 12|D_0{\bf \phi
}\pm eM\left( {\bf n}\times \pmb {\phi }\right) |^2  \nonumber \\
&&\pm M\left( J_0-egB\left( {\bf n}\times \pmb {\phi }\right) ^2\right)
-\frac 12g^2\left( {\bf n}\times \pmb {\phi }\right) ^2E^2+\frac geF_ik_i
\nonumber \\
&&\left. \mp gE_i\partial _i\left( {\bf n}\cdot \pmb {\phi }\right) +\frac
12\left( \nabla _1\pmb {\phi }{\bf \pm }\pmb {\phi }\times \nabla _2\pmb {%
\phi }\right) \pm \pmb {\phi }\cdot D_1\pmb {\phi }\times D_2\pmb {\phi }%
\right\} \mbox{ .}  \label{energy}
\end{eqnarray}
In order to make the system self-dual, we have chosen the potential as
\begin{equation}
U=\frac{e^2}{2G}\left[ (1-{\bf n}\cdot \pmb {\phi }{\bf )+}\frac \kappa
eM+g\left( {\bf n}\times \pmb {\phi }\right) ^2M\right] ^2+\frac
12e^2M^2\left( {\bf n}\times \pmb {\phi }\right) ^2  \label{sixteen}
\end{equation}
and we required that the following conditions must be satisfied
\begin{eqnarray}
B\mp \frac eG\left( (1-{\bf n}\cdot \pmb {\phi }{\bf )+}\frac \kappa
eM+g\left( {\bf n}\times \pmb {\phi }\right) ^2M\right) &=&0  \label{sd1} \\
E_i\pm \partial _iM &=&0  \label{sd2} \\
D_0\pmb {\phi }\pm eM\left( {\bf n}\times \pmb {\phi }\right) &=&0
\label{sd3} \\
\nabla _1\pmb {\phi }\mbox{\ }{\bf \pm }\mbox{\ }\pmb {\phi }\times \nabla _2%
\pmb {\phi } &=&0\mbox{ .}  \label{sd4}
\end{eqnarray}
Then we have a new form for the energy functional, namely
\begin{eqnarray}
\ &&{\cal E}=\int d^2x[E_i\left( \frac ge\varepsilon _{ij}k_j-g^2\left( {\bf %
n}\times \pmb {\phi }\right) ^2E_i\right) +\frac 12\partial _0M\partial _0M
\nonumber \\
&&\ -\frac 12g\partial _0M\partial _0({\bf n}\cdot \pmb {\phi })\pm M\left[
J_0+\kappa B+\partial _iE_i\right] \mp \partial _i\left( E_iM\right)
\nonumber \\
&&\ \mp g\left( E_i\pm \partial _iM\right) \partial _i({\bf n}\cdot \pmb {%
\phi })]+4\pi \int d^2xJ_{top}^0  \label{seventeen}
\end{eqnarray}
The last term in the expression above is just $4\pi Q_{top}$ , where $%
Q_{top} $ is the topological charge defined in eq.(\ref{topcharg}).

Using the equation (\ref{sd2}), integrating by parts and considering that
the surface terms goes to zero at infinity, we have
\begin{eqnarray*}
{\cal E} &=&4\pi \left| Q_{top}\right| \pm \int d^2x\left[ M\left(
J_0+\kappa B+\partial _iE_i+\frac ge\varepsilon _{ij}\partial _iJ_j\right)
+\frac 12\partial _0M\partial _0M\right. \\
&&\left. -\frac 12g\partial _0M\partial _0({\bf n}\cdot \pmb {\phi })\right]
\end{eqnarray*}
Now taking into account the Gauss law (\ref{gauss}) and considering static
solutions, clearly the energy functional is bounded by below as
\begin{equation}
{\cal E}\geq 4\pi \left| Q_{top}\right| {\ }  \label{bound}
\end{equation}
and this bound is saturated when the Bogomol'nyi equations (\ref{sd1}-\ref
{sd4}) are satisfied.

It is worth mentioning the role played by the neutral scalar field $M$. In
fact, it is essential to obtain a self-dual model, as was first pointed by
Lee {\it et al.} \cite{lee2} in the context of the
Maxwell-Chern-Simons-Higgs model. Still in this context, but with nonminimal
coupling, similar result was presented in refs. \cite{marc}.

The requirement for finite energy solutions restrict our boundary
conditions. From the potential (\ref{sixteen}) we see that the boundary
condition is
\begin{equation}
\lim_{r\rightarrow \infty }\pmb {\phi }=\pm {\bf {n}}\mbox{ .}
\label{limit2}
\end{equation}

\section{STATIC VORTEX SOLUTIONS}

Next, we solve the problem posed by the self-dual equations (\ref{sd1}-\ref
{sd4}). As we have seen, in the limit $g=g_c=-\frac e\kappa $ , the
equations of motion for the gauge field become first order equations.
Henceforth we restrict ourselves to this limit. From the $\alpha =0$
component of the equation (\ref{six}) and using (\ref{four}), we obtain the
following expression for the magnetic field $B$%
\begin{equation}
B=\pm \frac{eg_cM({\bf n}\times \pmb {\phi )}^2}{[1-g_c^2({\bf n}\times \pmb
{\phi )}^2]}\mbox{ .}  \label{B1}
\end{equation}
Note that we have used that $A_0=\mp M$, which can be see from (\ref{sd2})
for static solutions.

On the other hand, the Bogomol'nyi equations (\ref{sd1}-\ref{sd4}) provide
\begin{equation}
B=\mp \frac e{g_c}M\pm \frac{e(1-{\bf n}\cdot \pmb {\phi }{\bf )}}{%
1-g_c^2\left( {\bf n}\times \pmb {\phi }\right) ^2}\mbox{ .}
\end{equation}
Therefore, in this limit, the scalar field $M$ may be written in terms of
just the Higgs field, namely
\[
M=g_c\left( 1-{\bf n}\cdot \pmb {\phi }\right)
\]
and consequently
\begin{equation}
B=\pm eg_c\frac{({\bf n}\times \pmb {\phi )}^2}{1-g_c^2({\bf n}\times \pmb {%
\phi )}^2}\left( 1-{\bf n}\cdot \pmb {\phi }\right) \mbox{ .}  \label{B0}
\end{equation}
Now, the potential (\ref{sixteen}) can be rewritten as
\begin{equation}
U=\frac{e^2g_c^2}2\frac{(1-{\bf n}\cdot \pmb {\phi }{\bf )}^2}{\left[
1-g_c^2({\bf n}\times \pmb {\phi )}^2\right] }\left[ 1-({\bf n}\cdot \pmb {%
\phi )}^2\right] \mbox{ .}  \label{critpot}
\end{equation}
Note that in the limit $g_c\rightarrow 0,$ this potential becomes the one
considered in Ref. \cite{kimm}.

We search solutions which are invariant under simultaneous rotations and
reflections in space-time and target space. So, the spherically symmetric
fields are obtained by making use of the {\it ansatz }\cite{gladi1}
\begin{equation}
\pmb {\phi }(x)=(\sin f(r)\cos N\theta ,\sin f(r)\sin N\theta ,\cos f(r))
\label{ansatz}
\end{equation}
where $(r,\theta )$ are polar coordinates in the ${\bf x}$-plane, $N$ is a
non-zero integer and also defines the degree of a topological soliton ({\it %
vorticity }). To the gauge field given by ${\bf A}=N\frac{a(r)}{er}\widehat{%
\theta }$, the electric field vanishes, and the magnetic field is such that $%
B=N\frac{a^{\prime }}{er}$, where primes denote differentiation with respect
to $r$. Substitution of this {\it ansatz} into equations (\ref{sd3}), (\ref
{sd4}) and (\ref{B0}) gives
\begin{equation}
f^{\prime }(r)=\pm \frac Nr\frac{\sin f(r)}{1+g_c^2\sin ^2f(r)}\left[
1+a(r)\right] \mbox{ ,}  \label{f(r)}
\end{equation}
\begin{equation}
a^{\prime }(r)=\pm \frac rN\frac{g_c^2\sin ^2f(r)}{1-g_c^2\sin ^2f(r)}%
\;\left( 1-\cos f(r)\right) \mbox{ ,}  \label{a(r)}
\end{equation}
\begin{equation}
\frac 1eB=\pm g_c^2\frac{\sin ^2f(r)}{1-g_c^2\sin ^2f(r)}\left( 1-\cos
f(r)\right) \mbox{ .}  \label{B}
\end{equation}
In order to obtain solutions of the above equations it is necessary to
consider the asymptotic limit of the functions. The equations (\ref{f(r)})
and (\ref{a(r)}) can be decoupled and we will have a second order equation
for $f(r)$ given by
\begin{eqnarray}
f^{^{\prime \prime }}(r) &=&-\frac 1rf^{\prime }(r)+\frac{\left[ f^{\prime
}(r)\right] ^2}{\tan f(r)}-\frac{g_c^2\sin 2f(r)}{\left[ 1+g_c^2\sin
^2f(r)\right] ^2}\left[ f^{\prime }(r)\right] ^2  \nonumber \\
&&+\frac{g_c^2\sin ^3f(r)}{1-g_c^4\sin ^4f(r)}\left( 1-\cos f(r)\right)
\mbox{
.}  \label{f2}
\end{eqnarray}
Now we will find the solutions of the equations (\ref{f(r)}) and (\ref{a(r)}%
). First we consider the field variables near the origin. To ensure that the
field is nonsingular at origin we impose
\begin{equation}
f(0)=n\pi ,\mbox{ }n\in {\Bbb N}\mbox{ .}  \label{f0}
\end{equation}
Since ${\bf A}=N\frac{a(r)}{er}\widehat{\theta }$, clearly regular solutions
at origin require $a(0)=0.$ On the other hand, solutions for $f(r)$ are
symmetric about $f(r)=2\pi $. Therefore we can choose $f(0)=0$ and $f(0)=\pi
$. If we consider the latter condition it is interesting rewrite $f(r)$ as $%
f(r)=\pi +h(r)$. Considering also the lower sign and {\it positive} $N$, the
expression (\ref{f(r)}) for $h(r)\ll 1$ admits solution like
\begin{equation}
h(r)=C_0r^N\mbox{ .}  \label{h}
\end{equation}
Consequently
\begin{equation}
a(r)=-\frac 1{2N(2N+1)}g_c^2r^{4N+2}\mbox{ .}  \label{a0}
\end{equation}
If we consider the former boundary condition $f(0)=0$ and {\it negative} $N,$
near the origin, solutions of the eq. (\ref{f(r)}) are
\begin{equation}
f(r)=\overline{C}_0r^{-N}\mbox{ .}  \label{f00}
\end{equation}
Further, in this case, a solution for $a(r)$ may be written as
\begin{equation}
a(r)=-\frac{\overline{C}_0}{4N(1-2N)}g_c^2r^{-4N+2}\mbox{ .}  \label{apq}
\end{equation}
On the other hand, at infinity, two distinct asymptotic behaviours for the
solutions of (\ref{f2}) are possible. When $f(\infty )=\pi $ and $N$ is {\it %
positive}, again $f(r)$ can be written as $f(r)=\pi +h(r).$ In addition, an
asymptotic behaviour for $a(r\rightarrow \infty )$ compatible with finite
energy and eq. (\ref{a(r)}) requires $a(r\rightarrow \infty )\equiv -\alpha
_1.$ Finite solutions of the eq. (\ref{f2}) are of the form
\begin{equation}
h(r)=C_\infty r^{N(1-\alpha _1)}\mbox{ , }\alpha _1>1  \label{finfnt}
\end{equation}
and correspondingly
\begin{equation}
a(r)=-\frac{C_\infty ^2}{2N^2(1-\alpha _1)+2N}r^{2N(1-\alpha _1)+2}-\alpha _1%
\mbox{ .}  \label{agr1}
\end{equation}
Choosing the other boundary condition at infinity, namely, $f(\infty )=0$, $%
a(r\rightarrow \infty )\equiv \alpha _2$ and considering $N$ as a {\it %
negative }number{\it , }we have
\begin{equation}
h(r)=\overline{C}_\infty r^{N(1+\alpha _2)}\mbox{ , }\alpha _2>-1
\label{finft}
\end{equation}
and
\begin{equation}
a(r)=-\frac{\overline{C}_\infty ^4g_c^2}{2N}r^{4N(1+\alpha _2)+2}+\alpha _2%
\mbox{ .}  \label{agr2}
\end{equation}
Note that, in the above expressions, $\alpha _{1,2}$ are constants which
depend on our choice of topological or nontopological boundary conditions
and are to be considered just numerically determined.

For the ansatz (\ref{ansatz}) the topological charge can be expressed as
\cite{kimm}
\begin{equation}
Q_{top}=\frac N2\left[ cosf(0)-cosf(\infty )\right] \mp \frac{\alpha _{1,2}}%
2\left[ 1-cosf(\infty )\right] \mbox{ .}  \label{topcharge2}
\end{equation}
From the above equation one finds that solutions under conditions $f(0)=\pi $
and $f(\infty )=\pi $, have a non-integer topological charge which
caracterize a nontopological soliton \cite{ghosh1}. On the other hand,
boundary conditions $f(0)=0$ and $f(\infty )=\pi $ lead to solutions with $%
Q_{top}=N$ and so are called topological solitons. It is worth adding that
topological solitons have quantized energy ${\cal E}=4\pi \left| N\right| $
and not quantized magnetic flux $\Phi =\frac{2\pi }eN\alpha _2.$ In the case
of nontopological solitons, neither the energy ${\cal E}=4\pi \alpha _1$ nor
the magnetic flux $\Phi =\frac{2\pi }eN\alpha _1$ are quantized.

\section{NUMERICAL RESULTS}

The set of equations (\ref{f(r)}) and (\ref{a(r)}) was solved numerically
for the boundary conditions discussed above. First, we look for topological
solutions so we use the boundary conditions $f(0)=0$ and $f(\infty )=\pi $
with $N=-1.$ Nontopological solitons are obtained for $f(0)=\pi $ and $%
f(\infty )=\pi $ with $N=1.$ For constructing numerical solutions we have
used a shooting method. The asymptotic behaviours (\ref{h}), (\ref{a0}), (%
\ref{finfnt}) and (\ref{agr1}) were used for nontopological solutions while
expressions (\ref{f00}), (\ref{apq}), (\ref{finft}) and (\ref{agr2}) were
used for topological one. With regard to the latter case, the profiles for
the functions $f(r)$ and $a(r)$ for four different values of the critical
nonminimal coupling constant $g_c$ are given in Figs. 1-2 respectively,
whereas Figs. 4-5 show the functions $f(r)$ and $a(r)$ for the
nontopological case. The magnitude of the magnetic field $B$ as a function
of $r$ for $N=-1$ and $N=1$ is also presented in Figs. 3 and 6 respectively.
Here it is interesting to comment that for nontopological solitons (Fig. 6),
the magnetic field has doubly degenerate maxima, as first pointed out by
Ghosh and Ghosh \cite{ghosh1}. There degenerate and nondegenerate maxima
depend on the value of $a(\infty )$, while here this behaviour depends on
the values of $g_c,$ wherefore on the values of the relation $\frac e\kappa $
.

\section{CONCLUSIONS}

In this paper we have considered a gauged O(3) sigma model with both Maxwell
and Chern-Simons terms. Besides the minimal coupling, an extra term which
couples the complex scalar field directly to the gauge field strength, was
introduced. As in the context of Maxwell-Chern-Simons-Higgs systems the
presence of self-dual solutions is guaranteed only by introduction of a
neutral scalar field. Bogomol'nyi equations were constructed for a {\it {%
specific form}} of the potential and generic nonminimal coupling constant.
However, in contrast with several nonminimal models existing in the
literature, at a critical value for the nonminimal coupling parameter, we
have obtained topological as well as nontopological soliton solutions ( see
however \cite{marc}, where a dimensional reduction is used to construct a $%
N=2-D=3$ Maxwell-Chern-Simons-Higgs nonminimal model). Numerical solutions
for {\it both} topological and nontopological cases are obtained and
presented.

It is known that the critical condition $g_c=-\frac e\kappa $ turns a
Maxwell-Chern-Simons system in a pure Chern-Simons one \cite{stern,torres}.
So it is not surprising that for O(3) sigma models our results are similar
to the numerical solutions of Ghosh and Ghosh \cite{ghosh1} for an O(3)
sigma model where the gauge field dynamics is solely governed by a
Chern-Simons term. Indeed, for instance, the unusual behaviour of the
magnetic field for nontopological solitons with doubly degenerate maxima are
present in our model for specific values of $g_c$. To choose a value for $%
g_c $ means to choose a value for the coefficient $\kappa $ of the CS term,
so $\kappa $ can be used to define a class of solutions with degenerate or
non degenerate maxima of the magnetic field.

It is worthwhile to mention that, notwithstanding the resemblance commented
above, the Maxwell-Chern-Simons theories even in the large distance limit
are richer than the pure Chern-Simons model \cite{nemeth}. On the other hand
the moduli space dynamics of the solutions can be quite different due to the
contribution of the Maxwell term in the Lagrangian \cite{yee}.

Recalling the baby Skyrme model, the above discussed aspects of our model
may be used in an analysis of a relation between Maxwell, Hopf and Skyrme
terms in order to break the scale invariance of the sigma model.

\hspace{1.0in}

%TCIMACRO{\TeXButton{acknowledgments}{\centerline{\bf ACKNOWLEDGMENTS}}}
%BeginExpansion
\centerline{\bf ACKNOWLEDGMENTS}%
%EndExpansion

We would like to thanks the referee for his criticisms and suggestions. F.
S. A. Cavalcante and M. S. Cunha are supported by CAPES and FUNCAP,
respectively.

\newpage \ %\end{references}

\centerline{\bf FIGURE CAPTIONS}

Figure 1: A plot of $f(r)$ as a function of $r$ for N=-1 {\it topological}
soliton solutions for different values of the critical coupling parameter $g$
with $g=-0.8$ (dashed-dotted line), $g=-0.6$ (long-dashed line), $g=-0.4$
(dashed line), $g=-0.2$ (solid line).

Figure 2: A plot of $a(r)$ as a function of $r$ for N=-1 {\it topological}
soliton solutions for different values of the critical coupling parameter $g$
and the parameter $a(\infty )\equiv \alpha _2$ with $g=-0.8$ and $\alpha
_2=6.3590$ (dashed-dotted line ), $g=-0.6$ and $\alpha _2=3.7824$
(long-dashed line ), $g=-0.4$ and $\alpha _2=2.2078 $ (dashed line), $g=-0.2$
and $\alpha _2=1.0641$ (solid line).

Figure 3: The magnitude of the magnetic field $B$ as a function of $r$ for
N=-1 {\it topological} soliton solutions for different values of the
critical coupling parameter $g$ with $g=-0.8$ (dashed-dotted line), $g=-0.6$
(long-dashed line), $g=-0.4$ (dashed line), $g=-0.2$ (solid line).

Figure 4: A plot of $f(r)$ as a function of $r$ for N=1 {\it nontopological}
soliton solutions for different values of the critical coupling parameter $g$
with $g=-0.025$ (dashed-dotted line), $g=-0.02$ (long-dashed line), $%
g=-0.015 $ (dashed line), $g=-0.01$ (solid line).

Figure 5: A plot of $a(r)$ as a function of $r$ for N=1 {\it nontopological}
soliton solutions for different values of the critical coupling parameter $g$
and the parameter $a(\infty )\equiv -\alpha _1$ with $g=-0.025$ and $\alpha
_1=4.8459$ (dashed-dotted line), $g=-0.02$ and $\alpha _1=5.1790$
(long-dashed line), $g=-0.015$ and $\alpha _1=5.9420$ (dashed line), $%
g=-0.01 $ and $\alpha _1=9.3737$ (solid line).

Figure 6: The magnitude of the magnetic field $B$ as a function of $r$ for
N=1 {\it nontopological} soliton solutions for different values of the
critical coupling parameter $g$ with $g=-0.025 $ (dashed-dotted line), $%
g=-0.02$ (long-dashed line), $g=-0.015$ (dashed line), $g=-0.01$ (solid
line).

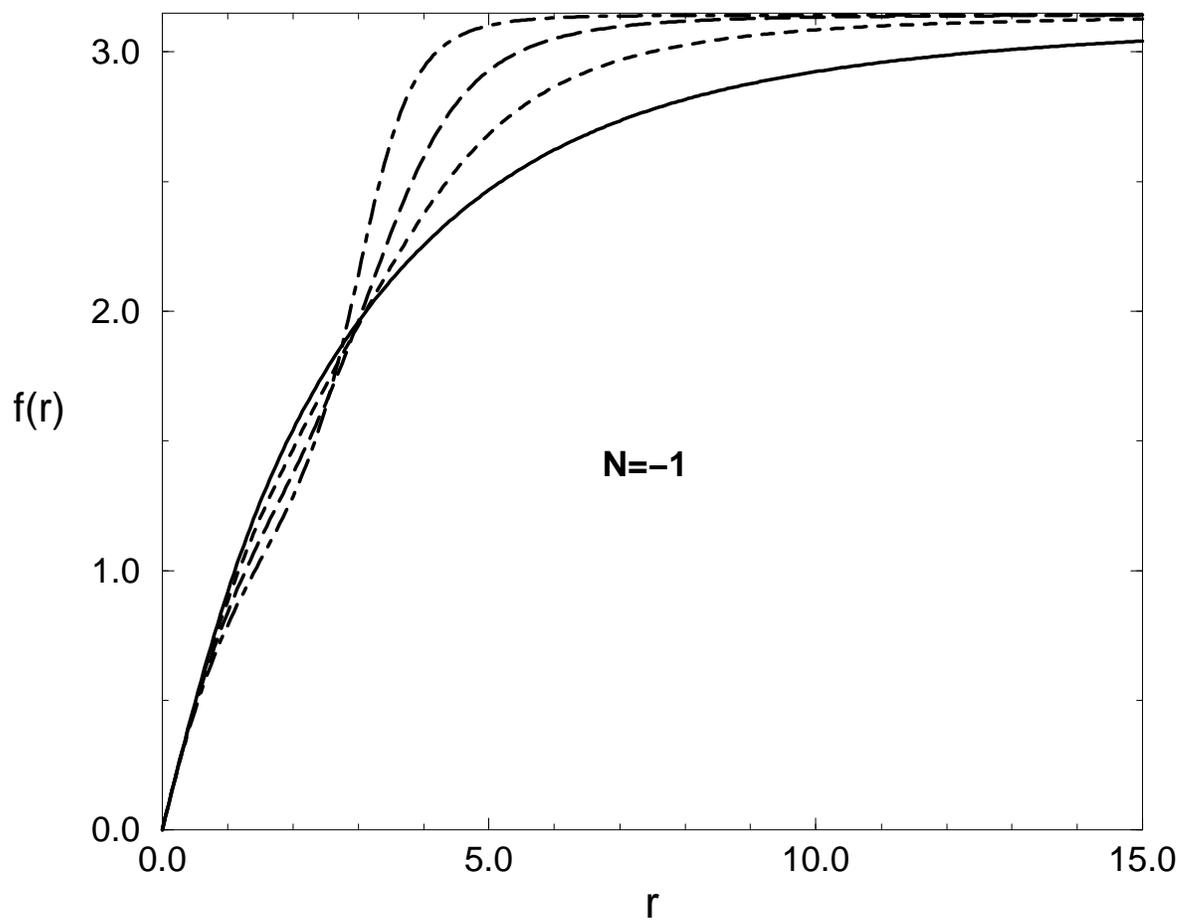
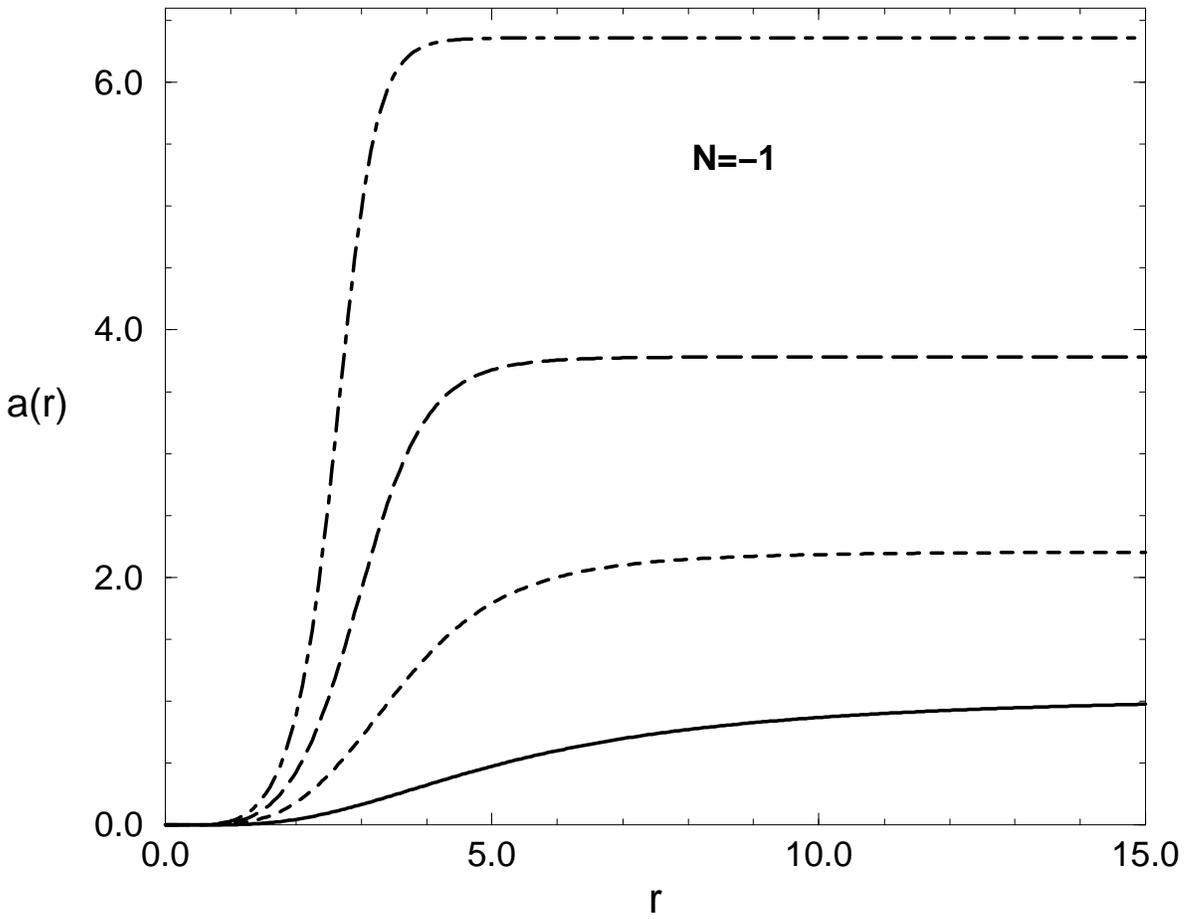
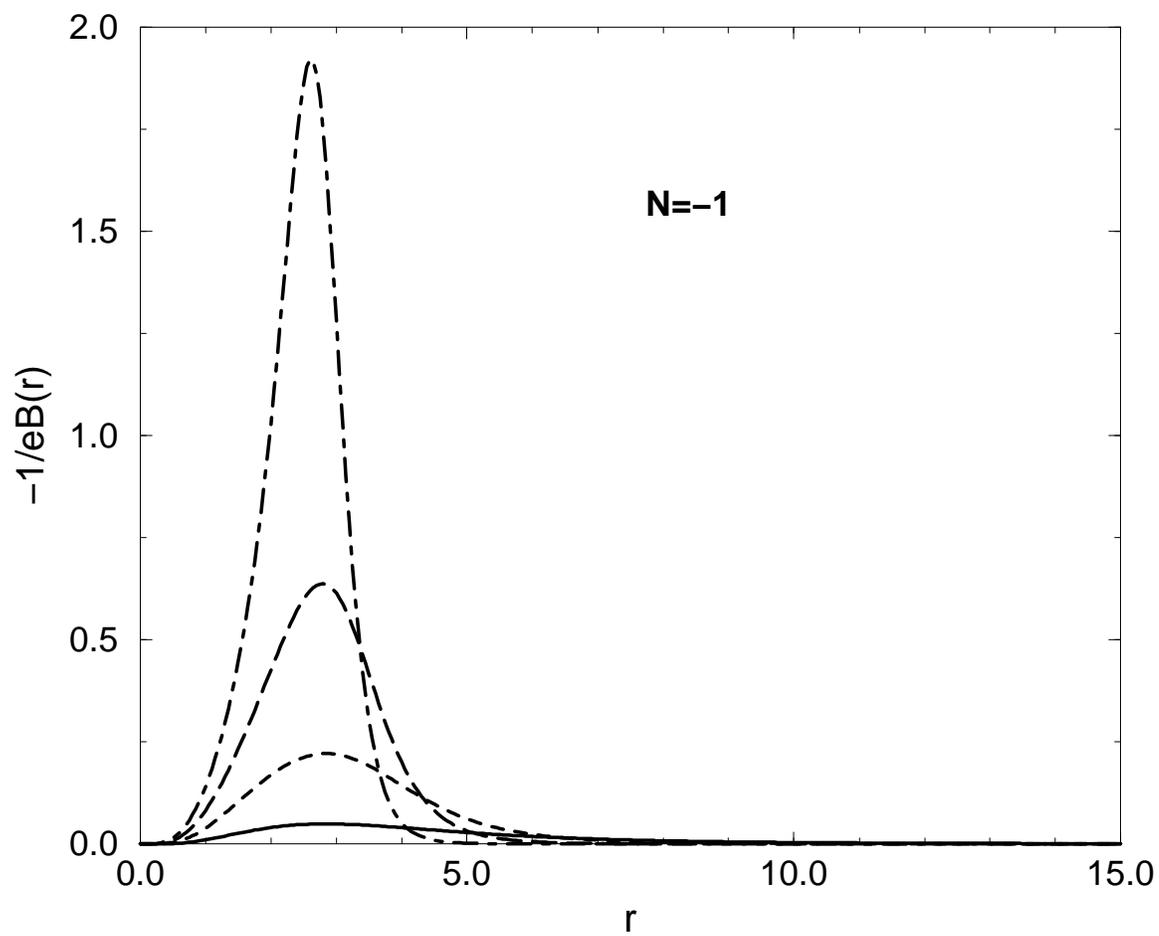
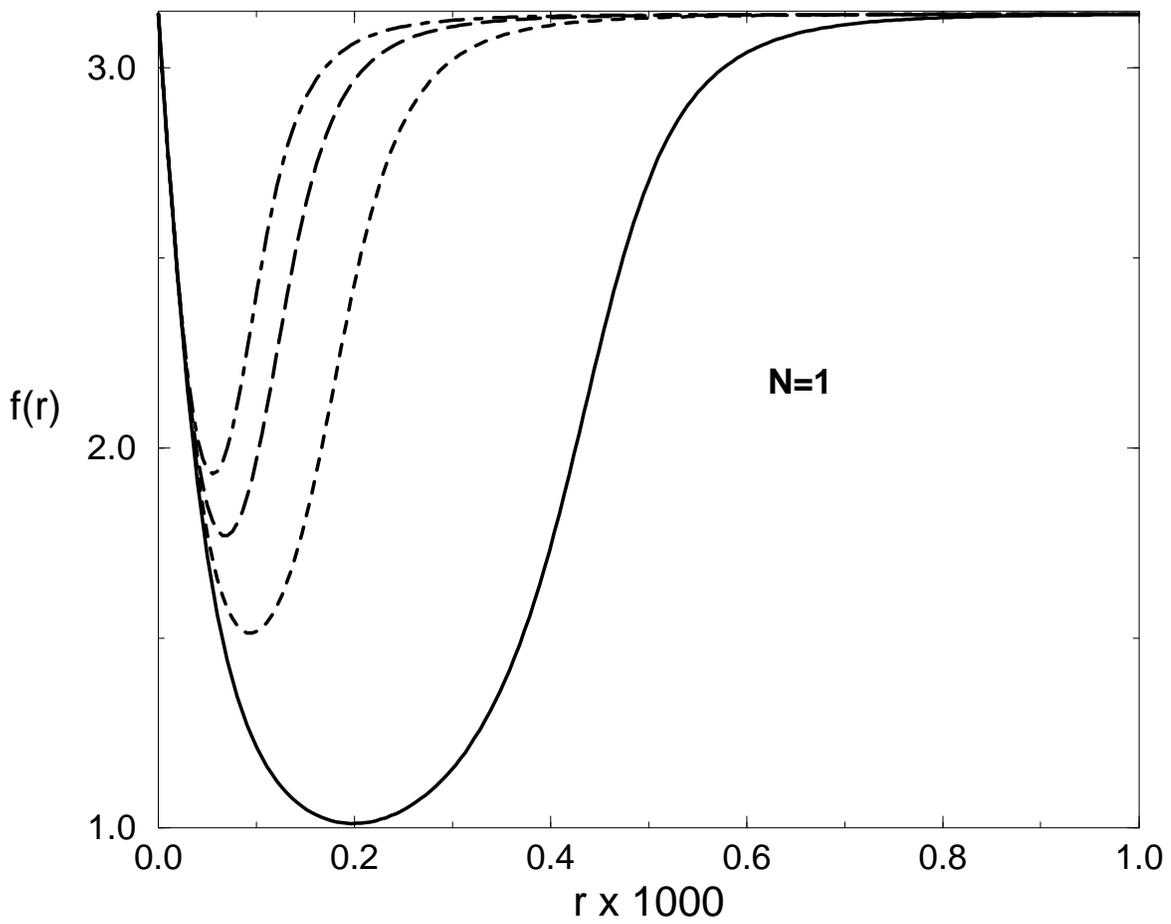
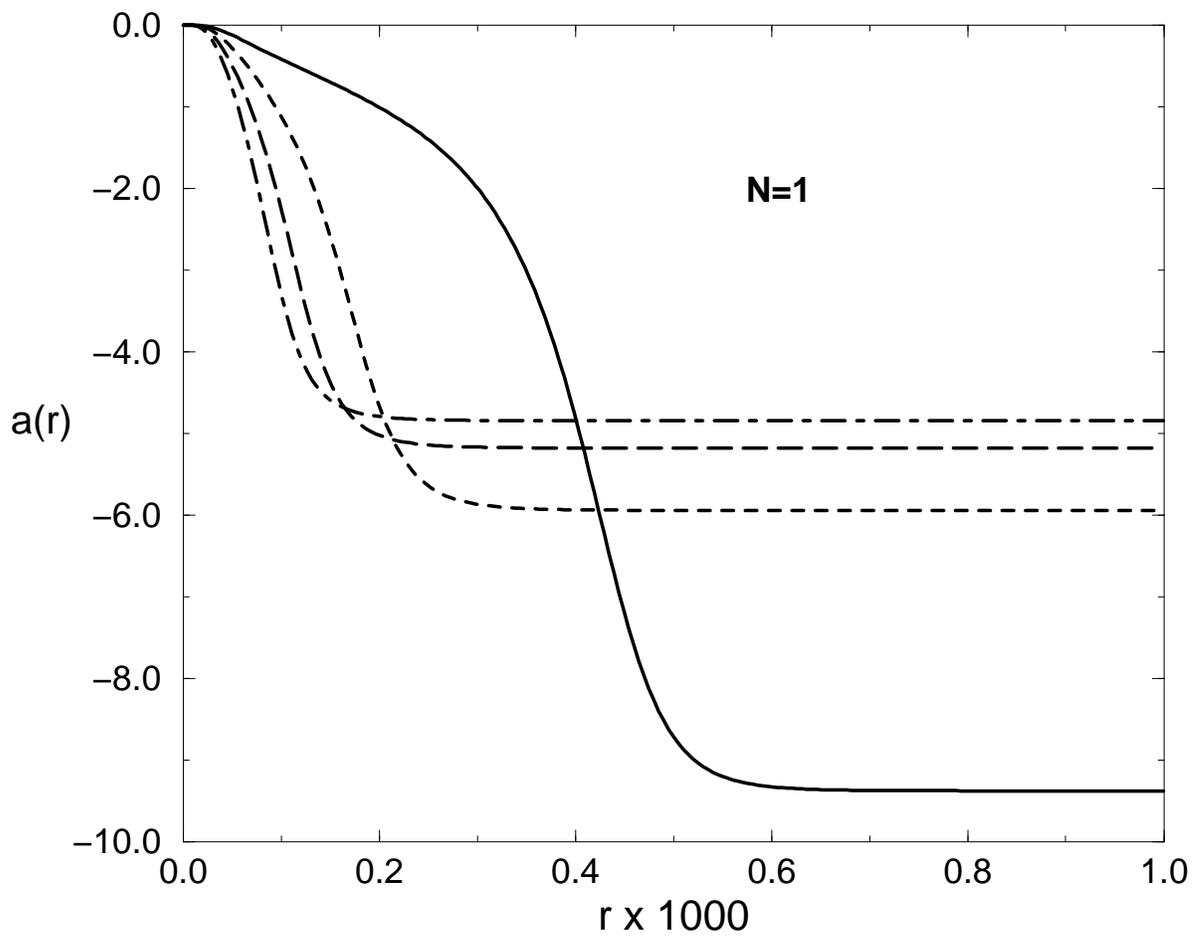
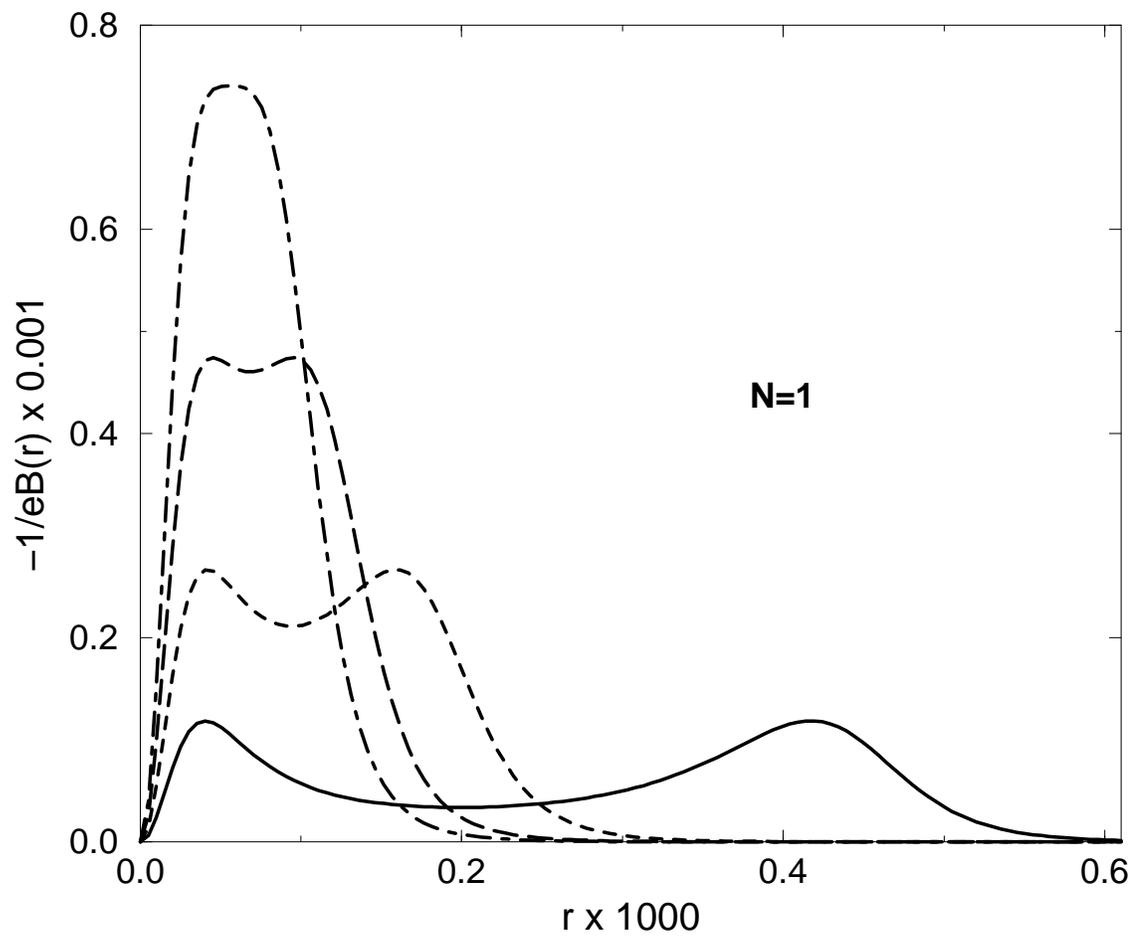
\end{document}